\documentclass[traditabstract]{aa} 
 
\usepackage{graphicx}
\usepackage{txfonts}
\usepackage{natbib}
\bibpunct{(}{)}{;}{a}{}{,}

\begin{document}

   \title{A comparison between observed Algol-type double stars in the Solar neighborhood and evolutionary computations of galactic case A binaries with a B-type primary at birth
         }

   \author{N. Mennekens
           \and D. Vanbeveren
          }

   \institute{Astronomy and Astrophysics Research Group, Vrije Universiteit Brussel, Pleinlaan 2, 1050 Brussels, Belgium\\
              \email{nmenneke@vub.ac.be, dvbevere@vub.ac.be}
             }

   \date{Received 25 November 2016 / Accepted January 2017}

  \abstract
   {We first discuss a large set of evolutionary calculations of close binaries with a B-type primary at birth and with a period such that the Roche lobe overflow starts during the core hydrogen burning phase of the primary (intermediate mass and massive case A binaries). The evolution of both components is followed simultaneously allowing us to check for the occurrence of contact binaries. We consider various models to treat a contact system and the influence of these models on the predicted Algol-type system population is investigated. We critically discuss the available observations of Algol-type binaries with a B-type primary at birth. Comparing these observations with the predictions allows us to put constraints on the contact star physics. We find that mass transfer in Algols is most probably not conservative, that contact during this phase does not necessarily lead to a merger, and that angular momentum loss must be moderate.
}

            
   \titlerunning{Algol-type double stars in the Solar neighborhood}
   
   \authorrunning{Mennekens \& Vanbeveren}

   \maketitle

\section{Introduction}

The evolution of binaries with a B-type primary at birth (i.e. a primary with a ZAMS mass between $\sim$3 M$_{\odot}$ and $\sim$20 M$_{\odot}$) has been the subject of numerous studies. Computations of galactic case B\footnote{Case B means that the period of the binary is such that Roche lobe overflow (further abbreviated as RLOF) will start while the primary (= the mass loser) is a post-core hydrogen burning/hydrogen shell burning star. The RLOF encountered in this paper is case Br (radiative), indicating that mass transfer starts at a time when the loser's outer layers are still radiative, ensuring a dynamically stable mass transfer.} binaries in this mass range have been presented by Iben \& Tutukov (1985), Van der Linden (1987), de Loore \& Vanbeveren (1995), Chen \& Han (2003). Case A\footnote{Case A means that the period of the binary is such that RLOF will start while the primary is a core hydrogen burning star.} binary evolution where the RLOF is treated in a conservative\footnote{Conservative means that it is assumed that during RLOF the total system mass and angular momentum are conserved.} way was discussed by Nelson \& Eggleton (2001). The latter paper relies on an extended library of binary evolutionary calculations but unfortunately this library has never been made available. In 2005 we computed our own library of evolutionary computations for conservative case A binaries with a B-type primary at birth (Van Rensbergen et al., 2005). In section 2 we first describe briefly some evolutionary details important for the scope of the present paper. Note already that the RLOF process in case A and B binaries is composed of two phases: a rapid phase where mass transfer occurs on the thermal (=Kelvin-Helmholtz) timescale followed by a slow mass transfer phase on the nuclear timescale of the mass loser. In a case A binary the latter is of the order of the core hydrogen burning timescale of the mass loser.

Algol-type binaries have been intensively studied observationally and theoretically since the beginning of the binary evolutionary era, the late sixties/early seventies and this resulted in the following overall Algol-type binary definition useful for evolutionary purposes:

\begin{itemize}
\item the less massive star fills its Roche lobe and is the most evolved star (the Algol-paradox).
\item	the Algol-phase is the slow phase of mass transfer (typical mass transfer rates $\sim$10$^{-5}$ - 10$^{-7}$ M$_{\odot}$/yr).
\end{itemize}

As noted above the evolutionary timescale of the slow phase of the RLOF in case A binaries is of the order of the core hydrogen burning lifetime of the loser and this is much larger than the slow phase timescale in a case B binary. It is therefore expected that most of the Algol-type binaries are case A binaries in the slow phase of RLOF.

The main purpose of the present paper is to compare evolutionary predictions of case A binaries with observed properties of Algol-type binaries with the intention to draw conclusions concerning the evolution during the RLOF of intermediate mass and massive case A binaries. The evolutionary computations are discussed in section 2. Section 3 describes the observed population of Algol-type binaries. Section 4 then deals with conclusions resulting from the comparison between prediction and observation.

\section{Evolutionary computations}

\subsection{The binary code}

The Brussels binary evolutionary code originates from the one developed by Paczynski (1967). In the latter only the evolution of the primary (= mass loser) was followed, but it contained a detailed calculation of the mass transfer rate during Roche lobe overflow (RLOF). Of primary importance is the fact that this code modeled the gravitational energy loss when mass leaves the star through the first Lagrangian point, energy loss that is responsible for the luminosity drop that is typical of the loser’s evolution during its RLOF phase. A more detailed description of how this is implemented in the code is given by Paczynski (1967). At present our code is a twin code that follows the evolution of both components simultaneously (the code has been described in detail in Vanbeveren et al., 1998 a, b). The opacities are taken from Iglesias et al. (1992), the nuclear reaction rates from Fowler et al. (1975). Semi-convection is treated according to the criterion of Schwarzshild \& Harm (1958) and convective core overshooting is included as described by Schaller et al. (1992).

The twin code follows the evolution of the mass gainer and therefore an accretion model is essential. The present paper deals with case A binary evolution. In these binaries the period is short enough for the gas stream that leaves the first Lagrangian point to hit the mass gainer directly (e.g., no Keplerian disk is formed) and it therefore seems appropriate in order to calculate the effect of the mass gain process on the structure and evolution of the gainer using the formalism of Neo et al. (1977). In this model it is assumed that the entropy of the infalling material is equal or larger than that of the gainer's outer layers, and that there is no rotational mixing. In this ``snowfall'' model the accreted matter is hence distributed evenly over the gainer's surface (see also Ulrich \& Burger, 1976; Flannery \& Ulrich, 1977; Kippenhahn \& Meyer-Hofmeister, 1977). 

\subsection{Evolutionary computations}

Our data set contains evolutionary tracks of binaries with initial primary mass M$_{10}$/M$_{\odot}$ = 3, 4, 5, 6, 7, 8, 9, 12, 15, and 17, initial mass ratio q$_0$ = M$_{20}$/M$_{10}$ = 0.4, 0.6 and 0.9, initial period P$_0$ (in days) = 1, 1.5, 2, 2.5, 3, 4, 5, 6. This is the grid of initial parameters used by van Rensbergen et al. (2008) (the detailed tracks can be found at http://vizier.cfa.harvard.edu/viz-bin/VizieR?-source=J/A+A/487/1129). The tracks illustrate the following evolutionary characteristics.

\begin{figure}[]
\centering
   \includegraphics[width=8.4cm]{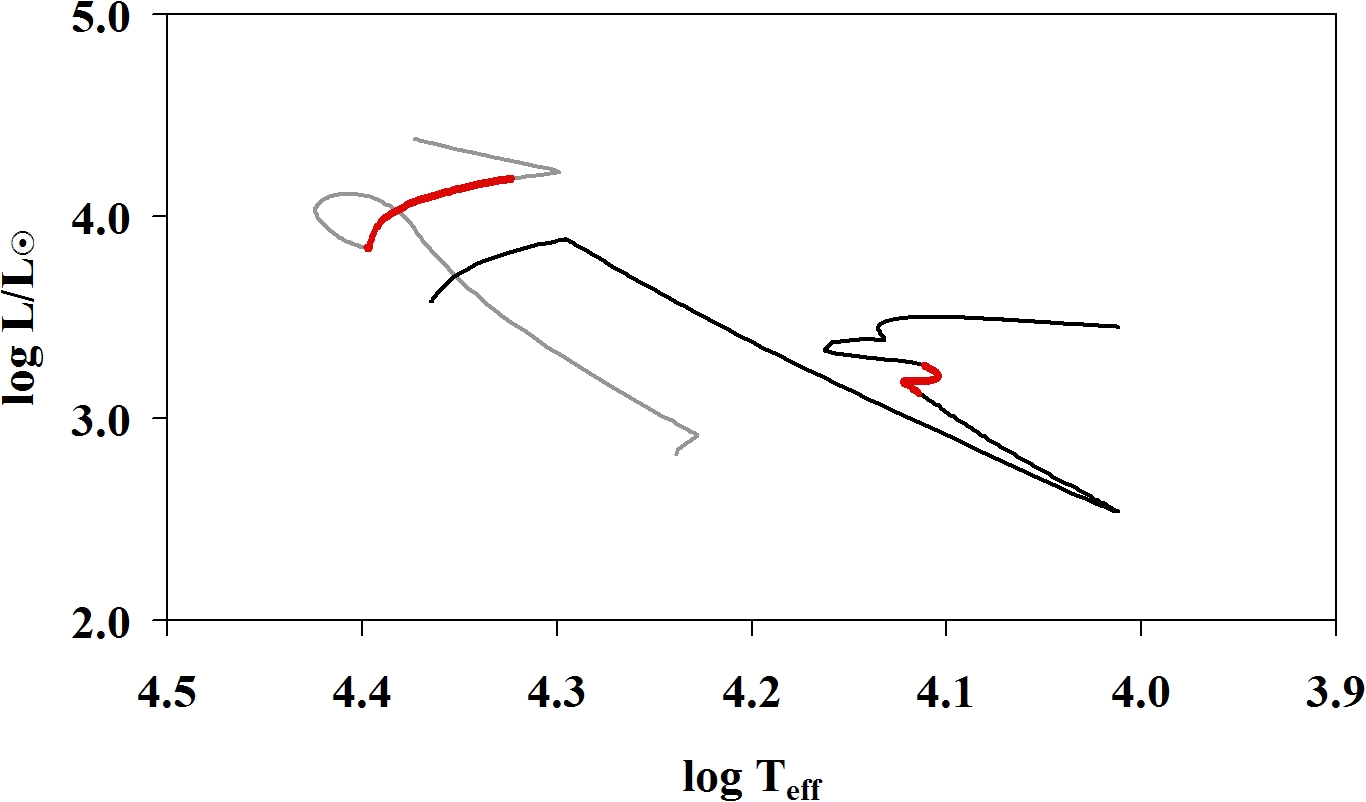}
     \caption{Typical run of tracks in the HRD of a binary star calculated with the Brussels binary evolution code. The shown tracks are for an initial 9.0+5.4 M$_{\odot}$ binary with an orbital period of 2.25 days. Black line is the track of the initially most massive star (the mass loser), gray line is the initially least massive star (the mass gainer). The slow Algol phase is highlighted in thick red.}
     \label{fig:1}
\end{figure}

\begin{figure}[]
\centering
   \includegraphics[width=8.4cm]{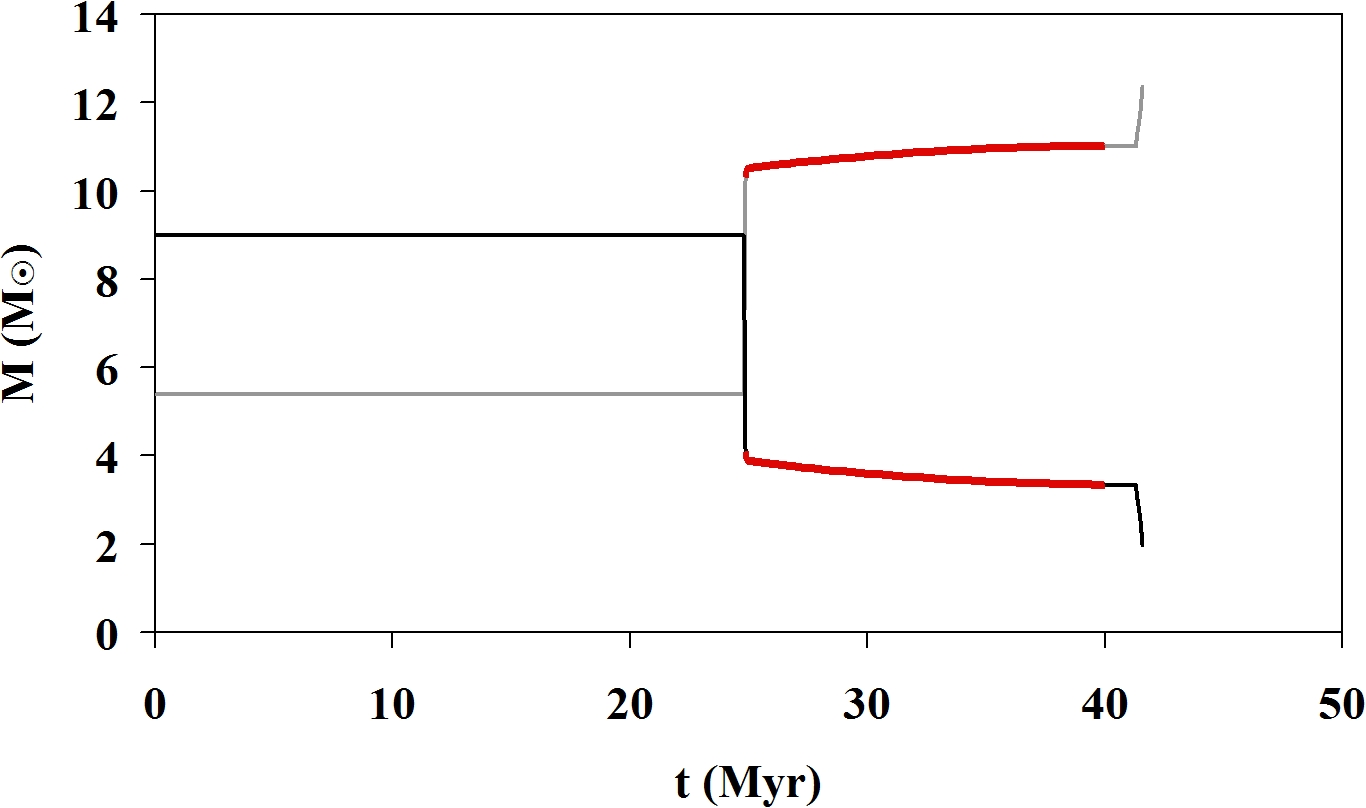}
     \caption{Typical temporal mass variation of a binary star calculated with the Brussels binary evolution code. The shown variations are for an initial 9.0+5.4 M$_{\odot}$ binary with an orbital period of 2.25 days. Black line is the mass variation of the initially most massive star (the mass loser), gray line is the initially least massive star (the mass gainer). The slow Algol phase is highlighted in thick red.}
     \label{fig:2}
\end{figure}

\begin{itemize}
\item	As already mentioned in the introduction the RLOF is characterized by a short phase where the primary loses mass at high rates (the rapid phase of RLOF acting on the thermal timescale of the mass loser), and a phase where the mass loss rate is moderate (small) that may continue during the whole remaining core hydrogen burning phase (the slow phase of RLOF acting on the nuclear timescale of the mass loser). Figure 1 illustrates the typical run of the tracks in the HR-diagram whereas Figure 2 shows the temporal variation of the mass.
\item	The behavior of the gainer during the rapid RLOF phase deserves some special attention. When the timescale of the rapid phase is smaller than the thermal timescale of the gainer the latter is driven out of thermal equilibrium and swells up. Depending on the initial mass ratio of the binary this may lead to the formation of a contact binary. It is \emph{unstable} contact in the sense that whenever the mass transfer stops the gainer very rapidly restores its thermal equilibrium, the star shrinks and the binary becomes semi-detached. In our data set contact happens in almost all the systems with initial mass ratio q$_0$ $\leq$ 0.6-0.7 corresponding to the subtype AR as defined by Nelson and Eggleton (2001). What really happens during AR contact is very uncertain. Possible suggestions are considered in the next paragraph. The evolutionary results of the systems in our data set were computed by ignoring eventual contact during the rapid RLOF phase. We will discuss the effect of this assumption.
\item	When the binary enters the slow phase of RLOF the gainer quickly restores its thermal equilibrium and follows a path in the HR-diagram that is entirely normal for its mass and chemical composition. 
\item	A contact phase quite different from the one discussed above occurs during the slow phase of RLOF in binaries with initial mass ratio close to unity. The mass gainer who regained its thermal equilibrium after the rapid RLOF follows a normal evolutionary path (core hydrogen burning followed by hydrogen shell burning). Depending on the initial period of the binary and on the present mass ratio the gainer may start filling its own Roche lobe forming a \emph{stable} contact system. When this happens we stop our calculations. Note that this corresponds to the binary types AS and AE introduced by Nelson and Eggleton (2001). 
\item	Of particular importance for the scope of the present paper is the conclusion that the overall stellar parameters of the loser at the end of the rapid phase of RLOF mainly depend on the exact moment during core hydrogen burning that the rapid RLOF starts and much less on the details of the RLOF process. The evolutionary computations discussed here assume conservative RLOF, but test calculations where for example 50\% of the mass lost by the loser is also lost from the binary, show that at the end of the rapid phase the final mass, the overall structure of the loser and its further evolution are very similar as in the conservative case.
\end{itemize}

\subsection{The evolution of contact systems during case A RLOF}

In the previous subsection we described two phases where contact sets in: contact during the rapid RLOF phase and contact during the slow RLOF phase. They are fundamentally different. When the latter happens it is conceivable that the binary merges and we stop the computations. The former one happens because the mass gainer accretes mass at a rate that is larger than the thermal rate. This causes a rapid swelling of the star and eventually it may fill its own Roche lobe. Our computations reveal that this type of contact happens in almost all our case A binaries with initial mass ratio $\leq$ 0.6-0.7, in correspondence with the detailed binary evolutionary models of Nelson and Eggleton (2001)\footnote{De Mink et al. (2007) studied case A evolution of binaries with subsolar metallicity and reached similar conclusions as for binaries with Solar metallicity.}. The evolution of contact binaries is still a matter of debate. It is one of the scopes of the present paper in order to explore the consequences of different assumptions related to the evolution of contact binaries for the prediction of the population of Algol-type binaries and to compare this with observations. 

Model 1: our calculations reveal that all case A binaries with initial mass ratio $\leq$ 0.6-0.7 enter contact during the rapid phase of RLOF. In model 1 we assume that these binaries merge and they therefore do not contribute to the Algol-binary population.

Model 2: contact during the rapid mass transfer phase is in most of the cases unstable meaning that when the rapid mass transfer stops the gainer rapidly shrinks and the system becomes semi-detached. In our second model we investigate the possibility that independent of contact the rapid RLOF remains conservative.

The subsequent models deal with the possibility that mass transfer during the rapid RLOF in case A binaries is accompanied by mass loss out of the binary\footnote{It is customary in order to describe the non-conservative RLOF by means of the parameter $\beta$ defined as the amount of mass lost by the loser that is effectively accreted by the gainer (0$\leq\beta\leq$1).}. We think that it is fair to state that despite many efforts the detailed physics of non-conservative RLOF is still poorly understood. There are however reasonable guesses. First, mass loss from a binary is obviously accompanied by orbital angular momentum loss. In the present paper we consider the two scenarios which are most frequently used by scientists studying binary populations. The first angular momentum loss formalism (AML1) assumes that when mass leaves the binary it takes with the specific orbital angular momentum of the gainer. The second (AML2): when mass leaves the binary it can do this via the second Lagrangian point (L$_2$) forming a circumbinary disk (van den Heuvel, 1993). The disk has to be sufficiently wide in order to avoid being fragmented by tidal forces. Following Soberman et al. (1997) we will assume that circumbinary disks are stable and do not have the tendency to fall back towards the binary if their radii are at least 2.3 times the binary separation. This disk model with disk radius equal to 2.3 times the binary separation defines AML2. 

It was shown in De Donder and Vanbeveren (2004) (see also Podsiadlowski et al., 1992) that AML2 yields a period variation given by

\begin{equation}
\frac{P_\mathrm{f}}{P_\mathrm{i}} = \left(\frac{M_{1\mathrm{f}}+M_{2\mathrm{f}}}{M_{1\mathrm{i}}+M_{2\mathrm{i}}}\right)\left(\frac{M_{1\mathrm{f}}}{M_{1\mathrm{i}}}\right)^{3\left[\sqrt{\eta}\left(1-\beta\right)-1\right]}\left(\frac{M_{2\mathrm{f}}}{M_{2\mathrm{i}}}\right)^{-3\left[\sqrt{\eta}\frac{1-\beta}{\beta}+1\right]},
\label{eq:P}
\end{equation}

with (following Soberman et al., 1997) $\eta = 2.3$. Moreover, Mennekens (2014) demonstrated that the period variation in case of AML1 can accurately be described by a similar formalism but with $\eta = 0.04$ which obviously implies a much smaller angular momentum loss than AML2.

Model 3: when contact happens during the rapid phase of RLOF all mass lost by the loser is lost from the binary and the angular momentum loss is described by AML1.

Model 4: similar to model 3 but with AML2\footnote{It may be interesting to note that Mennekens (2014) also investigated the case where the gainer rotates at its critical speed (a situation which is typical for case Br mass transfer) and where matter that leaves the binary takes with the specific orbital and rotational angular momentum of the gainer. He concluded that formalism (1) with $\eta = 0.9$ describes the resulting period variation accurately.}.

Contact during the rapid phase of the RLOF in case A binaries with mass ratio $\leq$ 0.6 is a shallow contact. But the smaller the mass ratio the deeper becomes the contact. It may therefore seem reasonable to adopt a non-conservative model where $\beta$ = 1 when q$_0$ $\geq$ 0.6 and decreasing when q$_0$ decreases. To illustrate we have chosen a linear decreasing function when q$_0$ $<$ 0.6 becoming zero at q$_0$ = 0.2.  

Model 5: a linear decreasing $\beta$-function and AML1.

Model 6: similar to model 5 but AML2.

Model 7: similar to model 5 but $\beta$ = 0.5 when q$_0$ $\geq$ 0.6 (hence $\beta$ linearly decreases from 0.5 to 0 as q$_0$ decreases from 0.6 to 0.2).

\section{The observed sample of Algol-type binaries}

\begin{figure}[]
\centering
   \includegraphics[width=8.4cm]{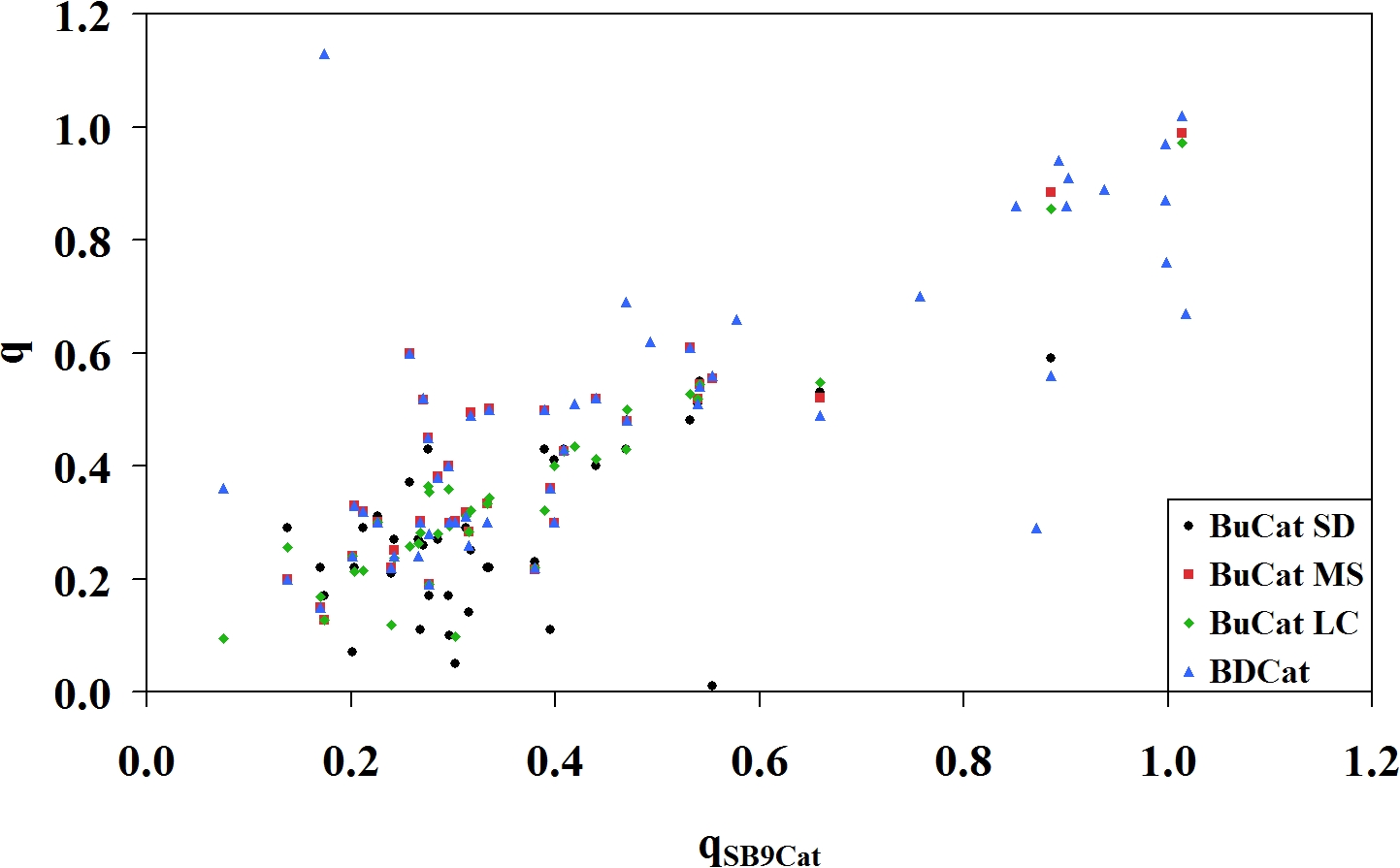}
     \caption{Comparison of binary mass ratios according to the SB9Cat (horizontal axis) with those according to several others (vertical axis) for 55 systems: black dots are BuCat SD, red squares are BuCat MS, green diamonds are BuCat LC and blue triangles are BDCat (see text for acronyms).}
     \label{fig:3}
\end{figure}

\begin{figure}[]
\centering
   \includegraphics[width=8.4cm]{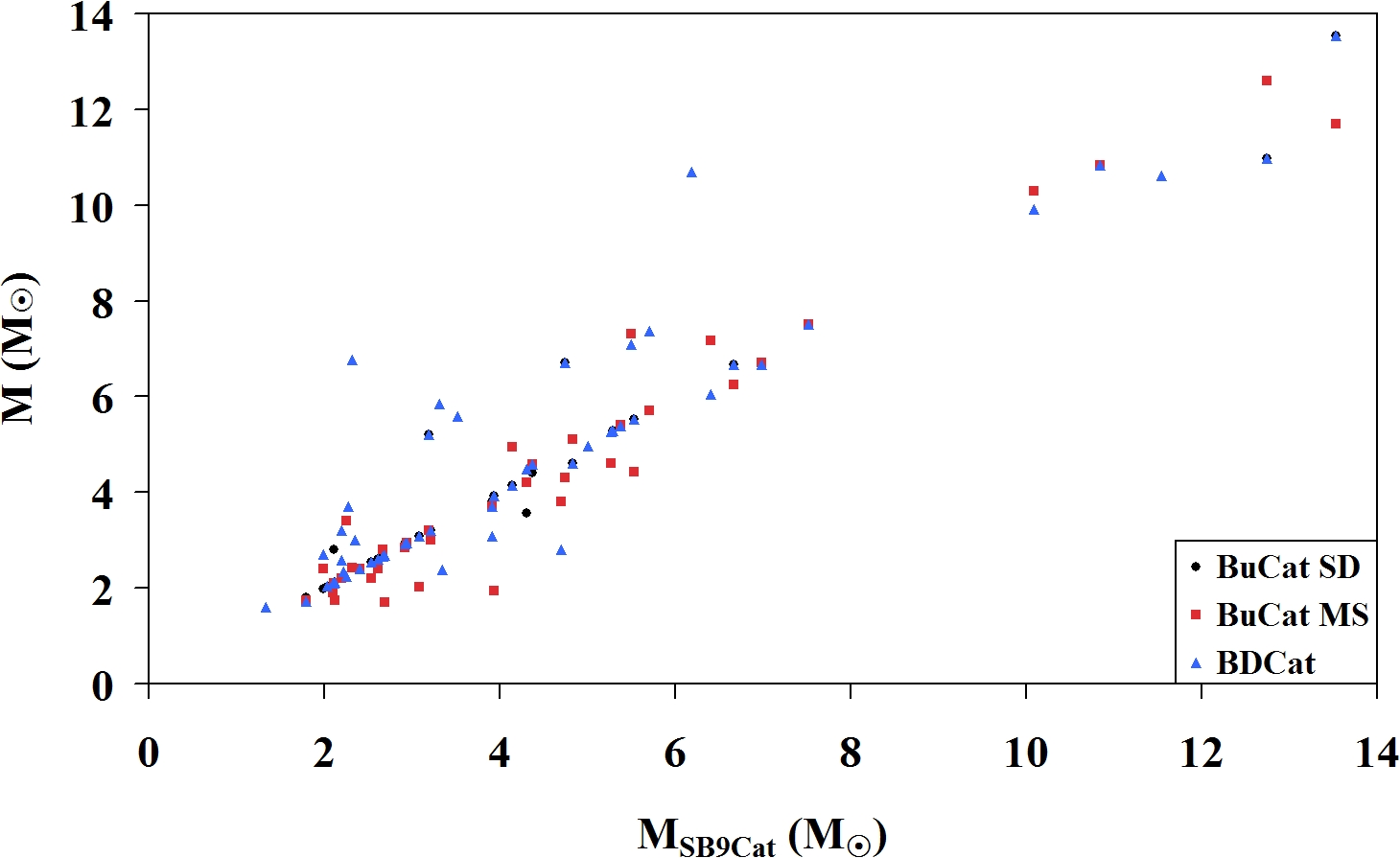}
     \caption{Comparison of gainer masses (in M$_{\odot}$) according to the SB9Cat (horizontal axis) with those according to several others (vertical axis) for 55 systems: black dots are BuCat SD, red squares are BuCat MS and blue triangles are BDCat (see text for acronyms).}
     \label{fig:4}
\end{figure}

\begin{table*}
\centering
\caption{Properties of the 26 selected Algol-type binaries}
\begin{tabular}{c | c c c c c c c c c c}
\hline
name	&	M (M$_{\odot}$)	&	q	&	P (d)	\\
\hline
AS Cam	&	2.51	&	0.757	&	3.431	\\
$\beta$ Per	&	1.49	&	0.380	&	2.867	\\
BP Vul	&	1.41	&	0.811	&	1.940	\\
DM Per	&	1.73	&	0.315	&	2.728	\\
EK Cep	&	1.12	&	0.554	&	4.428	\\
IQ Per	&	1.74	&	0.493	&	1.744	\\
KU Cyg	&	0.82	&	0.173	&	38.439	\\
$\lambda$ Tau	&	1.70	&	0.266	&	3.953	\\
MY Cyg	&	1.81	&	1.014	&	4.005	\\
RS Sgr	&	2.34	&	0.335	&	2.416	\\
RS Vul	&	1.37	&	0.313	&	4.478	\\
RV Lib	&	0.42	&	0.172	&	10.722	\\
RZ Cnc	&	0.54	&	0.170	&	21.643	\\
S Cnc	&	0.17	&	0.075	&	9.484	\\
SX Aur	&	5.46	&	0.542	&	1.210	\\
SX Cas	&	1.43	&	0.296	&	36.567	\\
SZ Cen	&	2.32	&	1.017	&	4.108	\\
TU Mon	&	2.70	&	0.212	&	5.049	\\
U Cep	&	2.84	&	0.659	&	2.493	\\
U CrB	&	1.35	&	0.285	&	3.452	\\
U Oph	&	4.52	&	0.903	&	1.677	\\
U Sge	&	1.90	&	0.333	&	3.381	\\
V1647 Sgr	&	1.98	&	0.901	&	3.283	\\
V539 Ara	&	5.27	&	0.852	&	3.169	\\
V909 Cyg	&	1.76	&	0.886	&	2.805	\\
Z Vul	&	2.20	&	0.409	&	2.455	\\
\end{tabular}
\label{tab:26Algols}
\end{table*}

The observed sample of Algol-type binaries resulting from intermediate mass (and massive) case A binaries has been drawn from the Algol-binary catalogues of Brancewicz and Dworak (1980, BDCat) and of Budding et al. (2004, BuCat) and the 9th Catalogue of Spectroscopic Binary Orbits of Pourbaix et al. (2004, SB9Cat). The latter is Web based and continuously updated. Many systems are SB1-systems and some binary parameters proposed in the catalogues may therefore be subject to substantial errors. Even SB2-binaries may be subjects to errors. Algol orbital periods are typically well-determined. Masses are less well known, as they are mostly determined from the mass function and the inclination. Even the mass ratio is very dependent on the way in which it is determined, e.g. by using the light curve only, or by assuming that the gainer obeys main sequence star relations. More information, and the procedure with which we calculate our best-determination of q, is given in van Rensbergen et al. (2008). The situation is illustrated in Figures 3 and 4. We selected the binaries for which data are given in the BDCat and/or BuCat and in the SB9Cat. For these common binaries Figure 3 compares the binary mass ratio of the binaries listed in BDCat and/or BuCat (vertical axis) with the mass ratio given in the SB9Cat (horizontal axis). The data of BuCat is further subdivided in SD (semi-detached assumption), MS (based on main sequence calibration) and LC (based on light curve). Figure 4 is similar to Figure 3 but we consider the mass of the mass gainer in the Algol-type binaries. Ideally all data points are expected to lay on the bisectors but as can be noticed, the differences between the catalogues for a significant fraction of Algol-type binaries is considerable, differences that may illustrate the uncertainties in the data. In Figures 5(O) and 6(O) we plot the systems of the different catalogues in the mass donor (M)-system mass ratio (q=mass donor/mass gainer) diagram and in the mass donor (M)-system period (P) diagram. It is important to realize that the occupation area of the observed systems in both diagrams predicted by the different catalogues is very similar, indicating that when we will compare the observational diagrams with the theoretically predicted ones this comparison may be independent from the possible errors illustrated in Figures 3 and 4. We also selected the Algol-type binaries (26) where the different catalogues agree upon the data possibly indicating that the uncertainties are smaller than for the other systems. They are listed in Table 1 and their location in Figures 5(O) and 6(O) is shown by the large black dots.

\section{The predicted Algol-type binary population compared to observations}

\begin{figure*}[]
\centering
   \includegraphics[width=16.8cm]{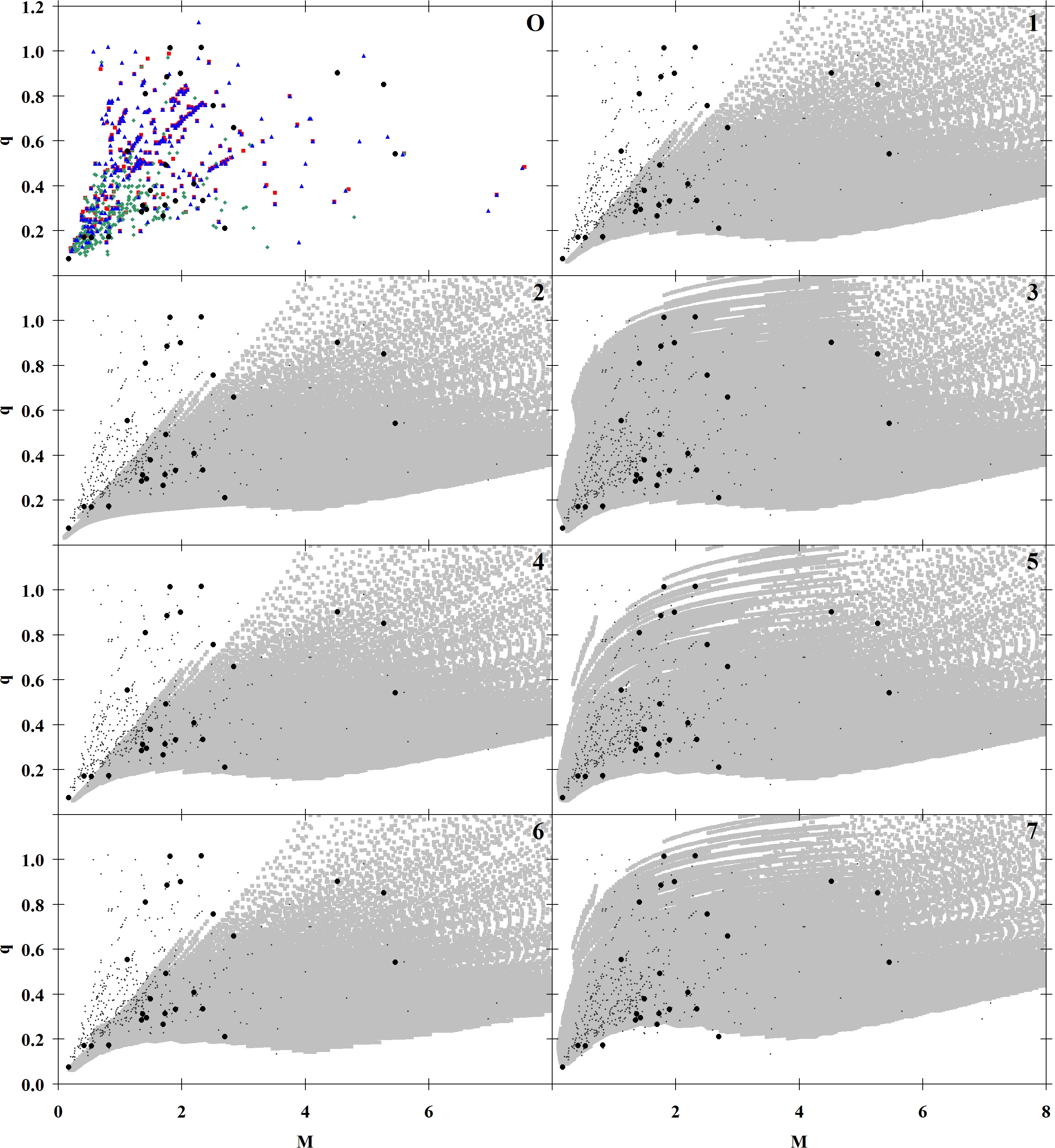}
     \caption{Populated zones in the donor mass (M) - mass ratio (q) parameter space. Panel O: 447 observations according to BuCat MS (small red squares), BuCat LC (small green diamonds) and BDCat (small blue triangles); large black dots are those for which the best-determination is also consistent with SB9Cat. Panels 1-7: theoretically populated zones with models 1 through 7, compared to all (small dots) and most reliable (large dots) observations.}
     \label{fig:5}
\end{figure*}

\begin{figure*}[]
\centering
   \includegraphics[width=16.8cm]{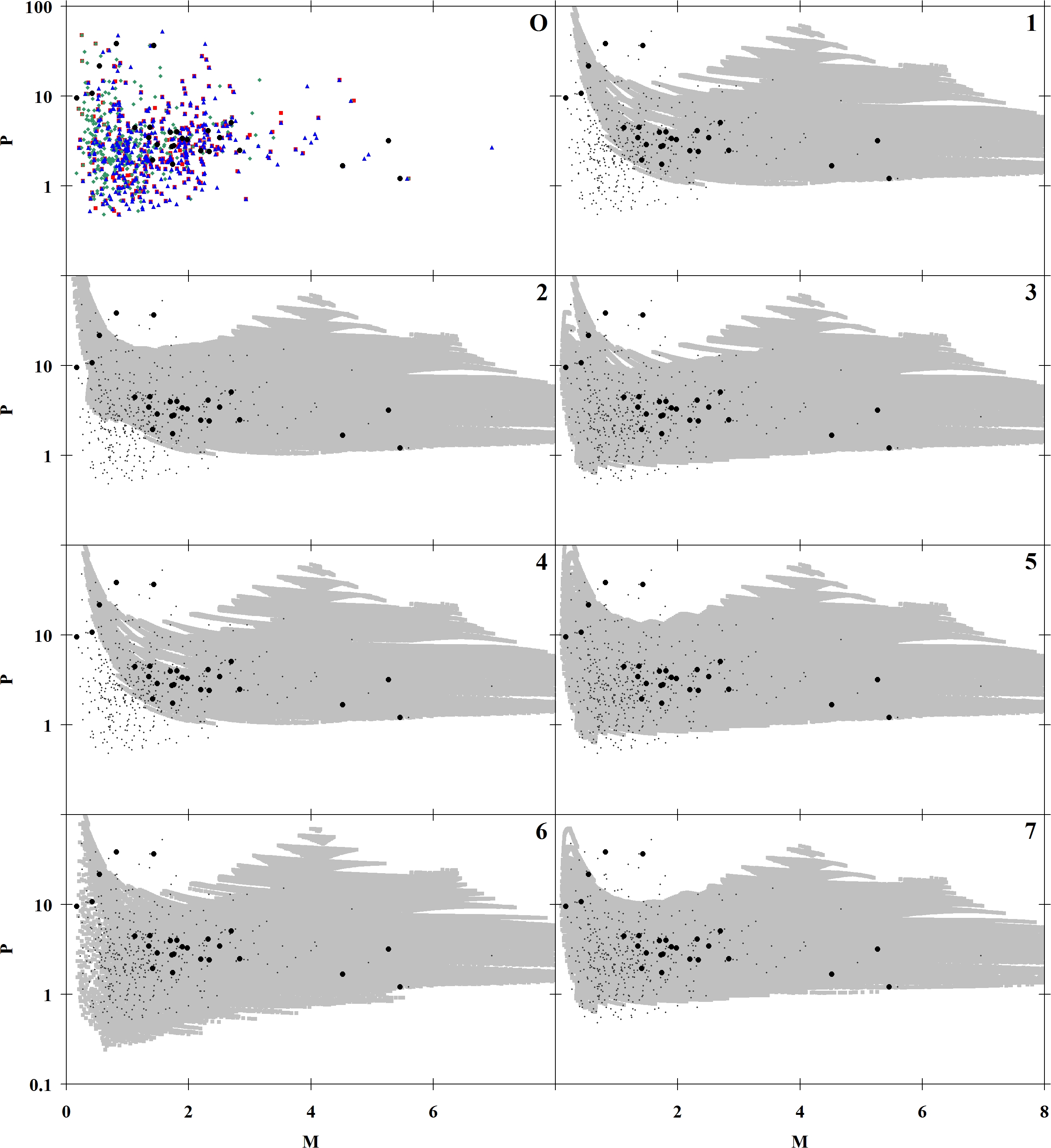}
     \caption{Populated zones in the donor mass (M) - orbital period (P) parameter space. Panel O: 447 observations according to BuCat MS (small red squares), BuCat LC (small green diamonds) and BDCat (small blue triangles); large black dots are those for which the best-determination is also consistent with SB9Cat. Panels 1-7: theoretically populated zones with models 1 through 7, compared to all (small dots) and most reliable (large dots) observations.}
     \label{fig:6}
\end{figure*}

Given a model listed in subsection 2.3 we determine the theoretically predicted occupation area in the M-q and M-P diagram and we compare with the observation diagrams (Figures 5 and 6). Our predictions are based on the evolutionary calculations and evolutionary properties discussed in section 2.2. Independent from the chosen model during the rapid phase of RLOF, when at the end of the rapid phase (when the gainer has reached thermal equilibrium) the two stars are still in contact, this is assumed to result in a merger.

In Figure 5(1)-5(7) we plot the region in the M-q diagram where we theoretically predict the presence of (intermediate mass and massive) Algol-type binaries for the 7 different models. The observations from Figure 5(O) are repeated in all panels for comparison purposes. Figure 6 is similar to Figure 5 but for the M-P region. We remind the interested reader that when an observed binary falls outside the theoretical region this means that this binary cannot be explained with the binary model that is considered. When this is the case for a significant number of observed systems we are inclined to reject the model. This leads to the following overall conclusions:

\begin{enumerate}
\item Models 1 and 2 do not fit the observations. We conclude that the rapid phase of case A RLOF is most probably not conservative whereas contact during the rapid phase of case A Roche lobe overflow does not necessarily lead to the merger of the binary.
\item Models 3, 5 and 7 cover the observations indicating that during the rapid phase of case A RLOF mass lost by the loser has to leave the binary. Comparison with models 4 and 6 (which are bad models) allows to conclude that when mass leaves the binary the accompanying loss of angular momentum must be moderate and of order of the specific angular momentum of the gainer. A large angular momentum loss (as in models 4 and 6) results in a binary merger explaining the similarity between model 1 and 4.
\item Most interestingly, the foregoing conclusions also apply for the 26 Algol systems of Table 1 which are considered as the Algols with the most certain orbital parameters.
\end{enumerate}

\section{Final remarks}

\begin{figure}[]
\centering
   \includegraphics[width=8.4cm]{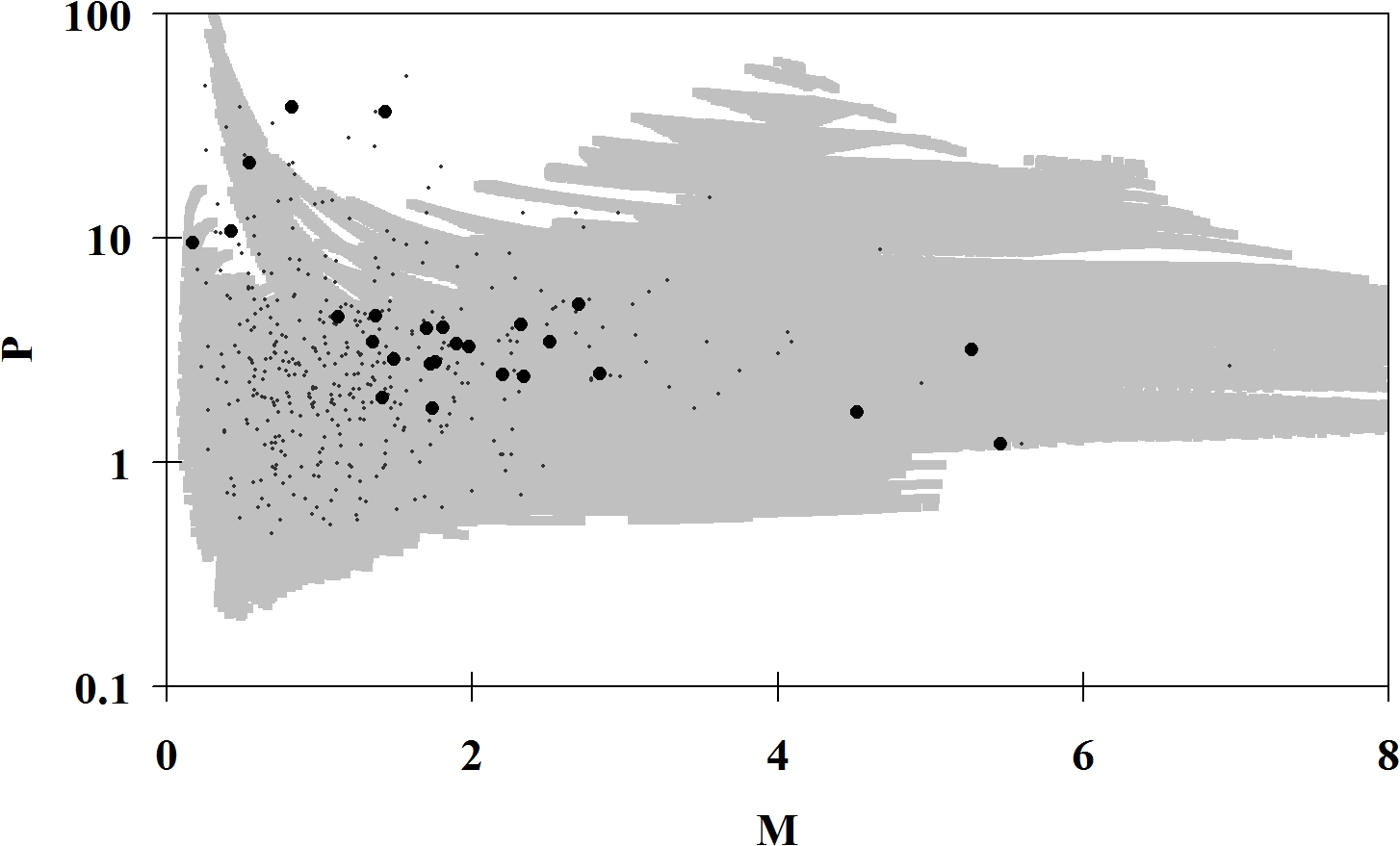}
     \caption{Same as Fig. 6(3), i.e. model 3, but for a hypothetical $\eta=0.125$ (see text).}
     \label{fig:7}
\end{figure}

\begin{figure}[]
\centering
   \includegraphics[width=8.4cm]{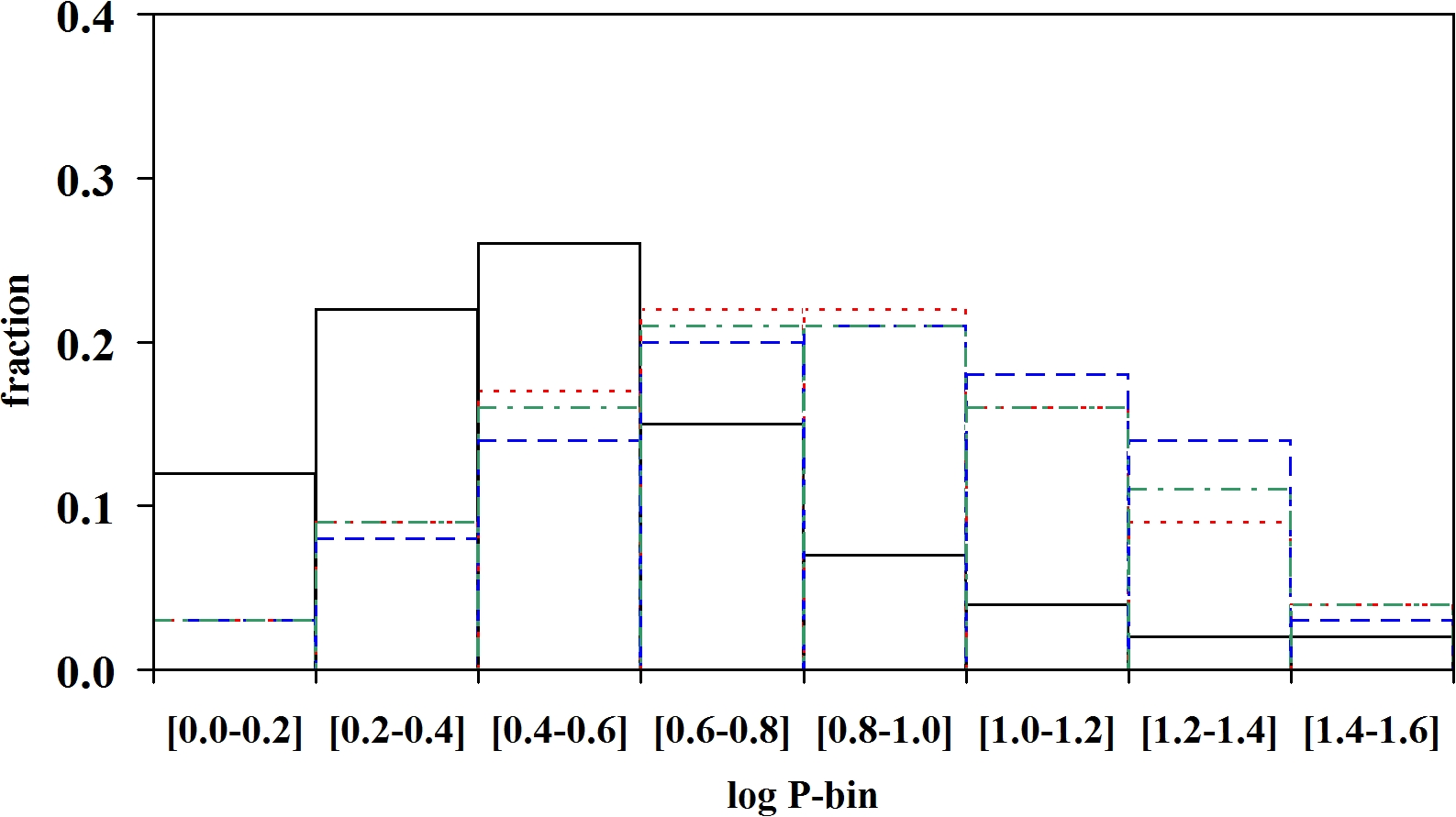}
	 \includegraphics[width=8.4cm]{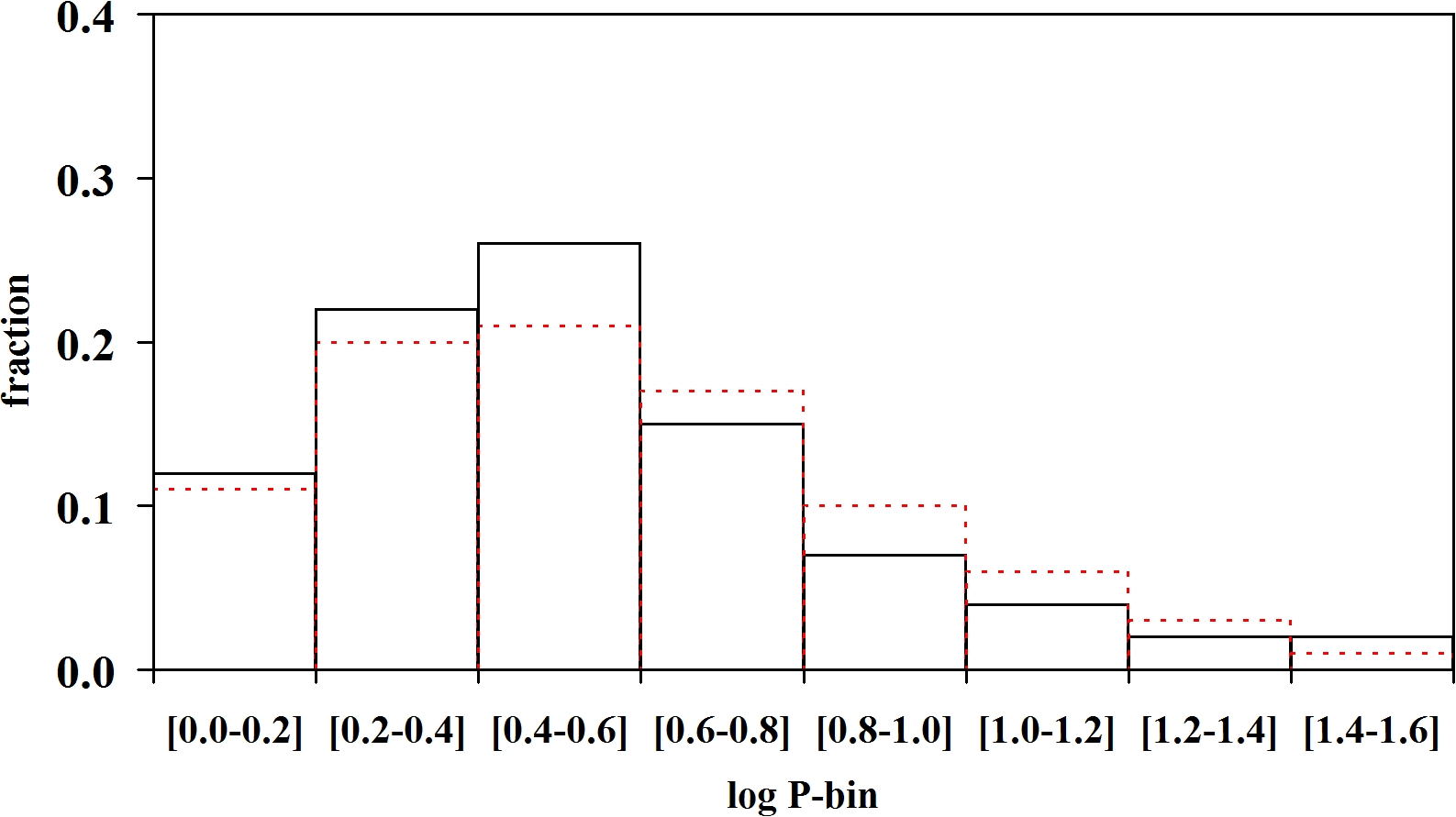}
     \caption{Top: number distribution of orbital periods for model 3 (dotted red), model 5 (dashed blue) and model 7 (dashed-dotted green), as well as observed distribution (solid black). Bottom: same (model 3 only), but for a hypothetical $\eta=0.125$ (see text).}
     \label{fig:8}
\end{figure}

\begin{figure}[]
\centering
   \includegraphics[width=8.4cm]{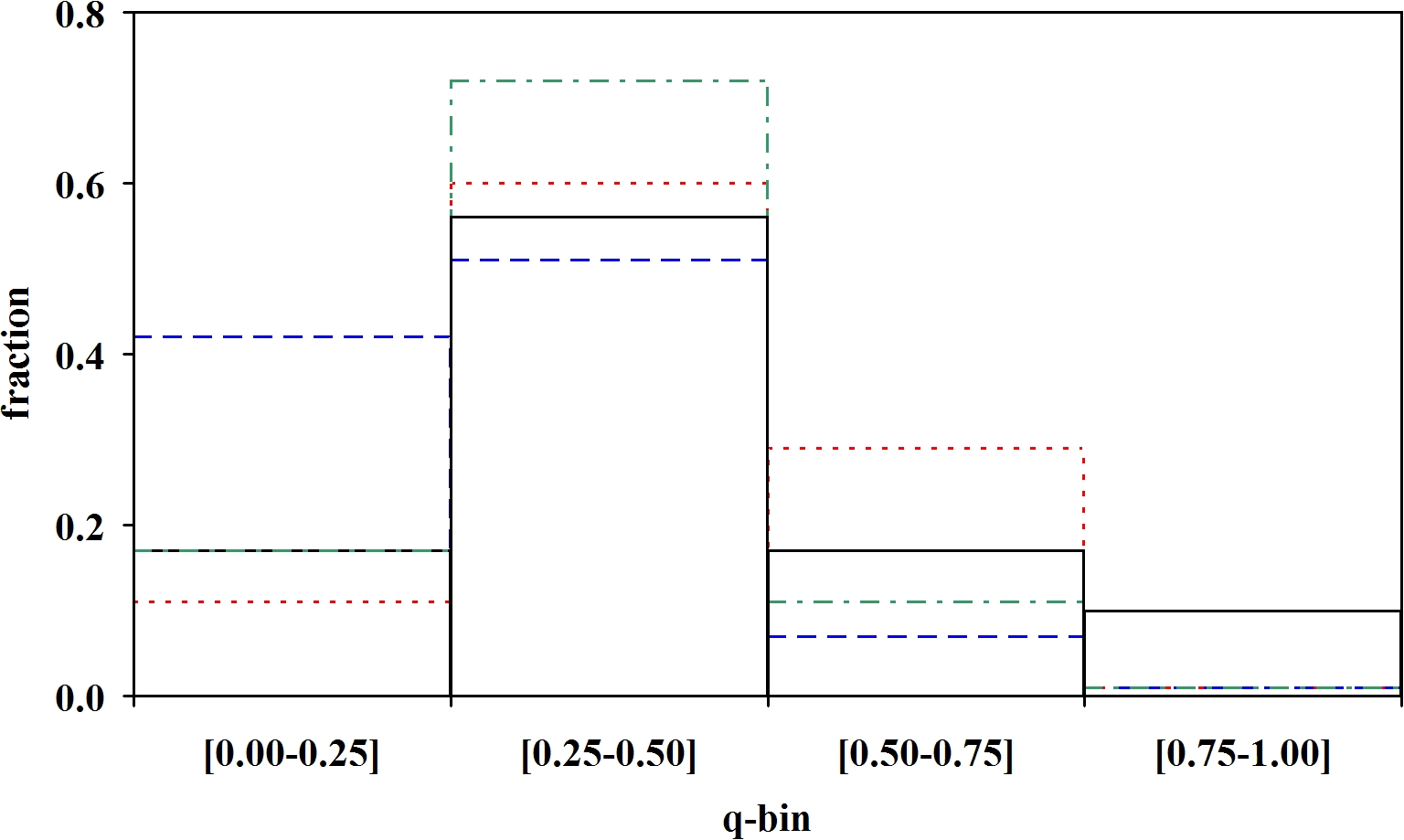}
     \caption{Number distribution of mass ratios for model 3 (dotted red), model 5 (dashed blue) and model 7 (dashed-dotted green), as well as observed distribution (solid black).}
     \label{fig:9}
\end{figure}

Our best models (3, 5 or 7) still have some difficulty in order to explain the smallest period Algols. As a thought experiment we therefore varied the value of $\eta$ in formalism (1). Figure 7 shows the resulting M-P diagram when $\eta=0.125$ which means that matter leaves the binary taking with angular momentum that is slightly larger than the specific orbital angular momentum of the gainer (corresponding to $\eta=0.04$, see section 2). As can be noticed the agreement is much better. Note that when mass leaves the binary via the mass gainer, in addition to the specific angular momentum of the gainer the leaving matter may also take with part of the specific rotational angular momentum of the star and therefore a slightly larger $\eta$-value is not unreasonable.

Notice that since we focused on the occupation areas in the M-q and M-P diagrams, our results and conclusions are independent from uncertainties typical for population number synthesis (PNS) simulations, like the initial mass ratio and period distributions of intermediate mass close binaries, the initial mass function of primaries of intermediate mass close binaries, and last but not least uncertainties in the Algol-lifetime of intermediate mass close binaries. Moreover, our conclusions do not critically depend on the observed q-uncertainties of Algol-type binaries discussed in section 2. Nevertheless, PNS compared to observed may reveal some interesting results. To illustrate, we focus on the q and period number distributions of the observed Algols. For the binary models considered in the present paper we calculated the predicted Algol q and P distributions by adopting a standard population of initial intermediate mass close binaries, with a Salpeter (1955) type initial mass function for the primaries, a Hogeveen (1992) initial q-distribution, a P-distribution that is flat in the Log and Algol-type lifetimes which result from our set of case A binary evolutionary tracks discussed in section 2.2. In Figure 8 we compare the predicted period distribution by adopting binary models 3, 5 and 7 (the best models after the occupation area analysis of section 4) with the observed period distribution\footnote{The observed period distribution represents the 303 Algol-type binaries listed in the catalogue of Brancewicz and Dworak (1980). Note that since the periods of most of the Algols are small, it can be expected that the period distribution is only marginally affected by observational bias.  This is illustrated by the fact that the period distribution resulting from other catalogues is very similar to the one that we show here.}. As can be noticed the agreement is rather poor. As argued above, the poor agreement may be due to the fact that the three models have some difficulty to explain the shortest period Algols and that this may be due to the adopted angular momentum loss when mass leaves the binary. To illustrate, Figure 8 also shows the model 3 simulation but with $\eta=0.125$ and as can be observed the correspondence is very satisfactory.

Figure 9 is similar to Figure 8 but for the q-distribution. The observed q-distribution has been discussed by van Rensbergen et al. (2008) and it should be kept in mind that compared to the observed P-distribution, it cannot be excluded that the q-distribution is much more affected by observational bias (see also section 3). Also here the binary models 3, 5 and 7 give by far the best correspondence with a preference for the models 3 and 7. A more thorough discussion on the effects on PNS simulations of uncertainties in e.g., the adopted initial distributions is beyond the scope of the present paper.

\begin{acknowledgements}
      We thank the referee Prof. Dr. Zhanwen Han for valuable comments and suggestions.
\end{acknowledgements}

\end{document}